\documentclass[review,12pt]{elsarticle}
\usepackage[a4paper, margin=2cm]{geometry}
\usepackage[english]{babel}
\usepackage[utf8]{inputenc}
\usepackage{amsmath}
\usepackage{amsfonts}
\usepackage{graphicx}
\usepackage{color}
\usepackage{comment}
\usepackage{hyperref}

\title{Discrete hierarchy of sizes and performances in the exchange-traded fund universe}
\date{\today}
\author[gent1,gent2]{B.~Vandermarliere}
\ead{Benjamin.Vandermarliere@UGent.be}
\author[gent2]{J.~Ryckebusch}
\ead{Jan.Ryckebusch@UGent.be}
\author[gent2]{K.~Schoors}
\ead{Koen.Schoors@UGent.be}
\author[eth1]{P.~Cauwels}
\author[eth1]{D.~Sornette}
\address[gent1]{Department of General Economics, Ghent University, Belgium}
\address[gent2]{Department of Physics and Astronomy, Ghent University, Belgium}
\address[eth1]{Department of Management, Technology and Economics, ETH Z\"{u}rich, Switzerland}

\begin{document}
\begin{abstract}
\noindent
Using detailed statistical analyses of the size distribution of a universe of equity exchange-traded funds (ETFs), we discover a discrete hierarchy of sizes, which imprints a log-periodic structure  on the probability distribution of ETF sizes that dominates the details of the asymptotic tail. This allows us to propose a classification of the studied universe of ETFs into seven size layers approximately organized according to a multiplicative ratio of 3.5 in their total market capitalization. Introducing a similarity metric generalising
the Herfindhal index, we find  that the largest ETFs exhibit a significantly stronger intra-layer and inter-layer similarity compared with the smaller ETFs.  Comparing the performance across the seven discerned ETF size layers, we find an inverse size effect, namely large ETFs perform significantly better than the small ones both in 2014 and 2015. 
\end{abstract}
\begin{keyword}
Econophysics \sep Exchange-traded funds (ETFs) \sep Probability density function of ETF sizes \sep Discrete scale invariance \sep ETF size layers and performance measures
\end{keyword}
\maketitle

\section{Introduction}
\label{Sec:introuduction}
An exchange-traded fund (ETF) can be thought of as a portfolio of  stocks, commodities, or bonds, which is traded like stocks on stock exchanges.  Exchange-traded funds have been made available as investment funds in the US in the early nineties and in Europe in the late nineties. Ever since, ETFs have emerged as a very important investment vehicle attracting ever increasing volumes of capital. Its attractiveness  is partly due to the relatively low management and transaction costs involved, an element that is particularly important in times of low yields and low interest rates. Exchange-traded funds represent an increasingly important investment vehicle with potential hazards for systemic risk and possible dangerous menaces for the financial system \cite{ETFRisk1} \cite{Bhattacharya2016} \cite{Malamud2015}. 
For example, it has been shown that arbitrageurs can contribute to cross-sectional return co-movement via ETF arbitrage. The presence of a stock in ETFs increases return co-movement at both the fund and the stock levels, where the effect is strongest among small and illiquid stocks \cite{da2013bellwether}. These days, ETFs come in many different types of flavours \cite{etfflavors}. For example, the degree of active management varies very much from one ETF to another.
 
The focus of this paper is on establishing a  taxonomy of the equity ETF landscape on the basis of their size. From our discussion we exclude leveraged ETFs and ETFs holding bonds and commodities, mainly to not overly complicate the analysis. As our focus is on determining the robust and stylized features of the equity ETF landscape using size, we do not segregate by types of ETFs, for example in terms of managed versus active versus passive, or index tracking ETFs.  

Size distributions often carry information about the underlying dynamics of a system. The analysis of the distribution of the equity ETF sizes  described below discloses some features that suggest departures from a simple power-like tail. The occurrence of a fat tail in the distribution of ETF sizes does not really come as a surprise given the well-documented approximate Zipf-law distribution of firm capitalisations \cite{Axtell2001}.  The fact, however, that there are strong indications that the tail is decorated with some log-periodic structure is remarkable. As this structure is connected with discrete scale invariance, one can infer some interesting constraints on the underlying dynamics of the equity ETF universe.  
Accordingly, we consider the disclosed log-periodic structure in the size distribution as a natural tool for classification of the universe of ETFs.   The inferred classification of the ETFs in several size layers is used to study various economic indicators. We address questions like: 'How similar are the various kinds of ETFs?'; 'How do ETFs distribute their holdings over the wide landscape of possible holdings?'; and 'Is there a connection between the ETF size and their performance?'.  These questions are naturally motivated by the existence of the size effect, exploited in the famous Fama-French  3 factor model \cite{FamaandFrench1993} that also addresses the fundamental issues of the relationships between diversification and performance.

The remainder of this paper is organized as follows. In Section~\ref{sec:analysisPDFofETFs} we present our empirical analysis of the equity ETF size distribution.  We start off (Section~\ref{yjuthgdqa}) with providing details of the ETF size data used and with performing a maximum-likelihood fit to their distribution. This reveals indications for an interesting  discrete hierarchical structure in the ETF size distribution that is discussed in more detail in Section~\ref{wrnhjki}.  In order to put this structure on more solid grounds and to get better hold on the disclosed periodicity in the size distribution, in Section~\ref{qettb} we pursue a detailed analysis of the ETF size distribution using kernel density estimation and Lomb periodograms.  In Section~\ref{subsec:mechanisms} we sketch some dynamical features of the ETF universe that may give rise to the observed hierarchical structure. We work out in detail how a model based on nonextensive (or, Tsallis) statistical mechanics, a current generalization of Boltzmann-Gibbs (BG) statistical mechanics, can give rise to the discerned oscillatory structures in the ETF size distribution. The basic premises of the proposed model is  that the system consisting of all ETFs operates as an open system in a capital reservoir. The size of the ETF system is subject to capital exchange with the reservoir, whereby there is a mechanism of both preferential attachment and growth. In Section~\ref{sec:significance} we introduce a classification into seven layers of the equity ETFs based on the discerned log-periodic hierarchy. We also explore how the economic properties vary over the various size layers. Thereby, we investigate the  intra-layer and inter-layer similarities (Section~\ref{subsec:intra_and_inter_similarity}), the variations in the stock holding ubiquity and capitalisation over the different layers  (Section~\ref{subsec:ubiquity_vs_cap}), and the connection between layer and performance (Section~\ref{subsec:performance_over_layers}). Our conclusions are drawn in Section~\ref{sec:conclusion}.
\begin{figure}[htb]
\centering
\includegraphics[scale=0.4, trim={0 0 0 0},clip]{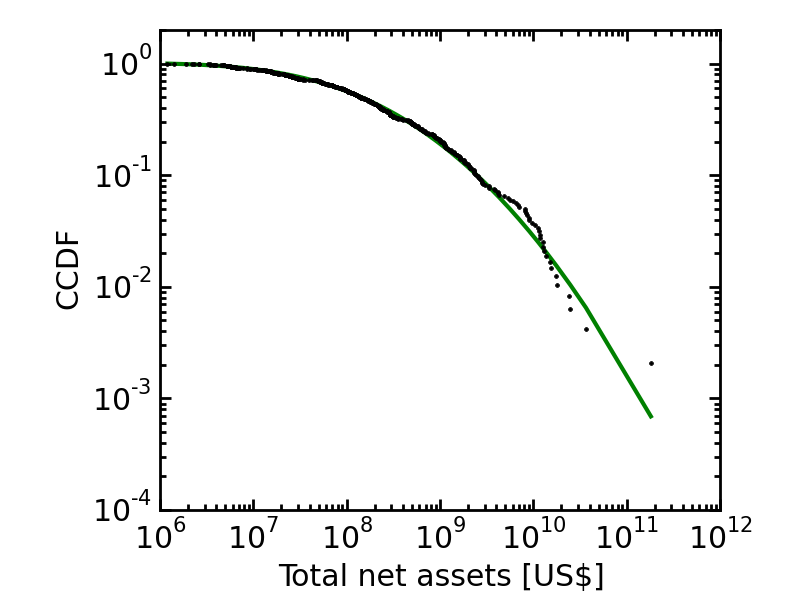}
\includegraphics[scale=0.4, trim={0 0 0 0},clip]{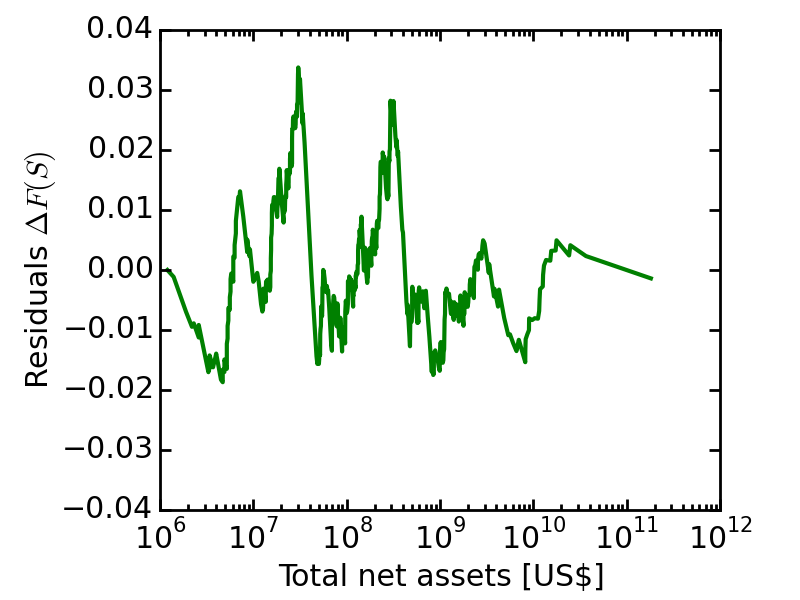}
\caption{The left figure shows the empirical complementary cumulative distribution function (CCDF)  of total net assets as a function of total net assets (decimal log-log scale) for the set of $479$ equity ETFs collected from Thomson Reuters Eikon in December~2014. The green full line is the maximum likelihood estimation of the lognormal distribution of Eq.~\ref{eq:lognormal} with $\widehat{\mu_{L}} = 18.7$ and $\widehat{\sigma_{L}} = 2.24$. The right figure shows the corresponding residuals -- the difference between the lognormal fit and the data -- as a function of total net assets.}
\label{fig:CCDFandLognormal}
\end{figure}
\section{Analysis of the distribution of ETF sizes}
\label{sec:analysisPDFofETFs}

\subsection{Distribution of total net asset values of ETFs \label{yjuthgdqa}}

At the end of $2014$, we collected data for all exchange-traded funds (ETFs) labelled as equity ETFs from Thomson Reuters Eikon. 
This resulted in a set of $479$ ETFs for which we obtained the total net assets and the entire composition of their portfolios. In total, this comprised $11,643$ different assets and about $100,000$ positions, for a total net assets over all ETFs of $1.399 \times 10^{12}$ US\$. Figure~\ref{fig:CCDFandLognormal} includes the complementary cumulative distribution function (CCDF) of the total net assets of ETFs, i.e., the fraction of ETFs of total net assets larger than or equal to $S$. Also shown is the CCDF of  the log-normal that best fits the data, as obtained by the maximum-likelihood method. The probability density function (PDF) of the log-normal law $\ln \mathcal{N}$ reads
\begin{equation}
\ln \mathcal{N} (\mu_{L}, \sigma_{L}^2) = \frac{1}
{x \; \sqrt[]{2 \pi \sigma_{L}^2}} e^{- \frac{(\ln x - \mu_{L} )^2}{2 \sigma_{L}^2} } \; ,
\label{eq:lognormal}
\end{equation}
with $\mu_{L}$ the location and $\sigma_{L}$ the scale parameter whose maximum-likelihood estimates are $\widehat{\mu_{L}} = 18.7$ and $\widehat{\sigma_{L}} = 2.24$. This corresponds to the mode (or most probable) ETF size of approximately 130$\times 10^6$ US\$
and a mean ETF size of 1.6$\times 10^9$ US\$. The much larger value of the mean compared to the mode reflects the existence of a very strong ``fat tail'' quantified by $\widehat{\sigma_{L}}$.

When referring to fat tails, it is often convenient to use power law distributions. The tail of a log-normal distribution with large variance (as found here) is difficult to distinguish from a power law distribution (see e.g.~Ref.~\cite{MalPisSorUMPU11}
and Section~4.1.3 of Ref.~\cite{Sornettebook04}). Indeed, visually, the tail of the empirical CCDF shown  in Fig.~\ref{fig:CCDFandLognormal} seems roughly compatible with an asymptotic power law with an exponent of about $1$ (Zipf's law). Such an approximate asymptotic Zipf's law has been documented for the distribution of firm sizes \cite{Axtell2001}. The fact that a similar approximate behaviour in the asymptotic tail is observed for the distribution of ETF sizes is not really a surprise as it can be expected from the presence of two joint and mutually reinforcing mechanisms. First, it is well known that the size of individual firms approximately obeys Zipf's law \cite{Axtell2001, Ramsden2000, Axtell2006, SimonandBonini1958, Marsili2005}. This result is robust \cite{Ijrl1977} and has been confirmed for different countries \cite{Ramsden2000} and  for several measures of firm size including number of employees,
profits, sales, value added, and market capitalizations. Therefore, randomly generated portfolios with weights roughly proportional to firm capitalisations will also have an asymptotic Zipf distribution 
in their tail, as a result of the generalized central limit theorem (see Section 4 of Ref.~\cite{Sornettebook04} for a pedagogical presentation). Second, Zipf's law appears quite generically from the combination of three very robust ingredients, namely ETFs are born, they grow via proportional growth and then can also die or close. As outlined in Refs.~\cite{SaiMalSor09} (Chapter 10) and \cite{LeraSor16}, mergers and acquisitions do not change significantly the overall picture. If the stochastic component of proportional growth is large, Zipf's law is generically an excellent approximation of the tail \cite{SaiMalSor09,MalSaiSor13}.

\begin{figure}[htb]
\centering
\includegraphics[scale=0.4, trim={0 0 0 0},clip]{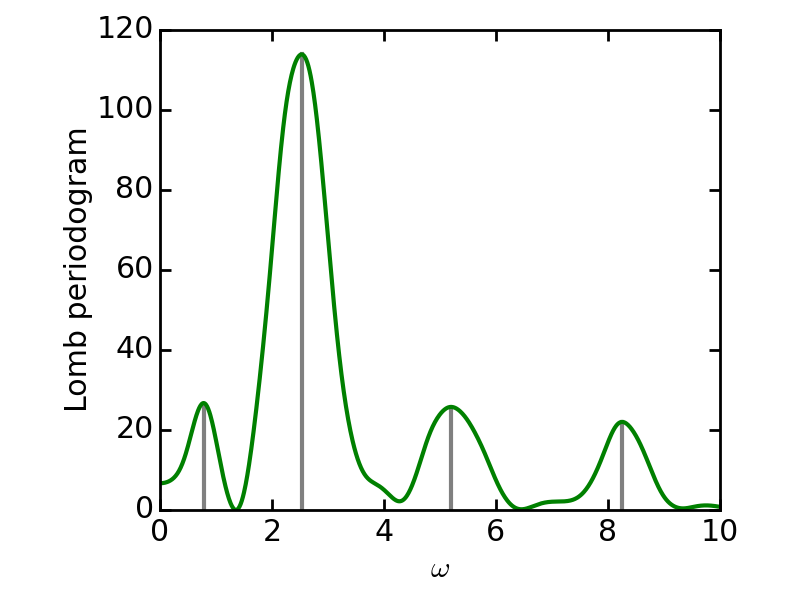}
\caption{Lomb periodogram of the residual function $\Delta F(S)$ shown in the right figure of Fig.~\ref{fig:CCDFandLognormal}. Here, $\omega$ is the conjugate variable to the logarithm of the ETF sizes. 
The occurrence of the three peaks at $\omega_1=2.5$ (large peak), $\omega_2=5.2 \approx 2 \omega_1$ and $\omega_3=8.2 \approx 3 \omega_1$ is interpreted in the text. As explained in the text, the peak at $\omega = 0.78$ is likely due to the conjunction of noise in the presence of a finite range of analysis.}
\label{fig:CCDFandLognormal_periodogram}
\end{figure}

Therefore, the observation of a fat tail that looks roughly like Zipf's law is not of much significance. What is much more surprising is the existence of very large deviations from a smooth tail, as made apparent by the structure of the residuals $\Delta F(S)$ of the lognormal calibration also shown in  Fig.~\ref{fig:CCDFandLognormal}. The pattern of these residuals clearly dominates the question of what is the asymptotic behaviour at large ETF sizes. A first preliminary conclusion is that there appears to be significant more texture to the tail of the CCDF
than just a power law or log-normal tail.  We now turn to the detailed quantitative analysis of these residuals.

\subsection{Evidence of a discrete hierarchical texture in the distribution of ETF 
sizes by spectral analysis of the residuals \label{wrnhjki}}

A visual inspection of the residuals $\Delta F(S)$ shown in Fig.~\ref{fig:CCDFandLognormal} suggests a noisy oscillation. To ascertain the significance of this observation, we calculate the Lomb periodogram of these residuals, shown in Fig.~\ref{fig:CCDFandLognormal_periodogram}. The use of the Lomb periodogram, instead of a Fourier transform, is required as a result of the non-even spacing of the pseudo-time variable, namely the logarithm of the total net assets $S$. Recall that the Lomb periodogram is a method for spectral analysis, which quantifies the contribution of each frequency to a given signal, based on the local least square fit of sine functions to the data \cite{scargle1982studies}.  In our case, the signal is the function $\Delta F (S)$ shown in the right panel of Fig.~\ref{fig:CCDFandLognormal} expressed as a function of $\ln S$. A statistically significant oscillatory component would mean that $\Delta F (S) $ can be expressed as 
\begin{equation}
\Delta F(S) = A + B \cos[\omega \ln S   +\phi] + \mathcal{O}\left( (\ln S)^2 \right)~, 
\label{rnwrgbvqd}
\end{equation}
where $(A, B,\phi)$ are three constants and $\mathcal{O}\left( (\ln S)^2 \right)$ is a second-order residual function of amplitude much smaller than $B$.

It is important to note that $\omega$ is not an angular frequency in the usual sense, as it is the conjugate variable to $\ln S$ and not to $S$. In other words, as already mentioned, the Lomb spectral analysis is performed in terms of the variable $\ln S$. Thus, the presence of periodicity in the $\ln S$ variable means that the residual function $\Delta F(S)$ is log-periodic in the function $S$, i.e.~it exhibits the symmetry of ``discrete scale invariance'' \cite{SornetteDSI98,ISI:000348403600019}.
In particular, $\omega$ is dimensionless. Fig.~\ref{fig:CCDFandLognormal_periodogram} exhibits an extremely large peak at $\omega = 2.5 \pm 0.2$, which  embodies the value of the scaling ratio 
$p_1:= \exp (2\pi / \omega_1 ) = 12.3$ for $\omega_1 = 2.5$ and quantifies the ratio of the geometrical series $S_n$ at which
the cosine in expression (\ref{rnwrgbvqd}) is equal to $1$ (i.e.  $\omega \ln(S_n) +\phi = 2\pi n$, where $n$ is an arbitrary integer).   According to extensive simulations in the possible presence of heavy-tailed and correlated noise \cite{ZhouSornetteLomb02}, one can ascertain that this peak at $\omega = 2.5 \pm 0.2$ is statistically highly significant. It expresses the existence of a discrete hierarchy of ETF sizes, roughly spaced  according to the ratio $p_1=12.3$. Note also the existence of the two smaller peaks at $\omega_2=5.2 \pm 0.2 \approx 2 \omega_1 $ and $\omega_3=8.2 \pm 0.4 \approx 3 \omega_1$. The presence of these harmonics strengthens the evidence for log-periodicity
\cite{ZhouSornetteTurb02,ZhouSornettePisTur03}.
The peak at the lowest value $\omega = 0.78$ corresponds to an oscillation of about the size of the entire range of values, which can be expected just from  cumulative noise effect \cite{Huangetalarti00} and we thus ignore it.

\subsection{Generalized derivative and Lomb periodogram of the PDF of ETF sizes \label{qettb}}
\label{subsec:Lombperiodogram}

In science, and especially in statistics, it is challenging to prove the absolute reality of an empirical observation. But one can scrutinise the data with a variety of distinct and complementary methods, which altogether may provide confirming evidence of the claimed phenomenon and thus stronger trust in its genuine existence. Because the claim of discrete scale invariance and of a discrete hierarchical structure in the distribution of ETF sizes is rather unexpected and of possible economic importance, we present a detailed analysis of the observed log-periodicity using a completely different methodology, which follows precisely the procedure described in \cite{Zhouetalfac3Dun05,fuchs2014fractal}.
The procedure has three components: (i) the kernel density estimation (KDE) of the probability density function (PDF)
(instead of using the CCDF) of the ETF sizes; (ii) the construction of the generalized $(H,q)$-derivative of the PDF, and (iii) the calculation of its Lomb periodogram.

\begin{figure}[htb]
\centering
\includegraphics[scale=0.4, trim={0 0cm 0 0},clip]{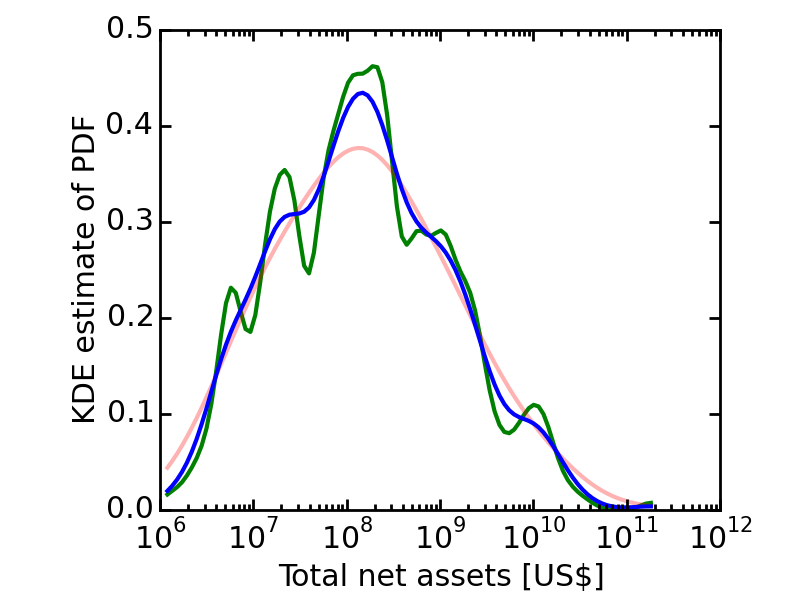} 
\includegraphics[scale=0.4, trim={0 0cm 0 0cm},clip]{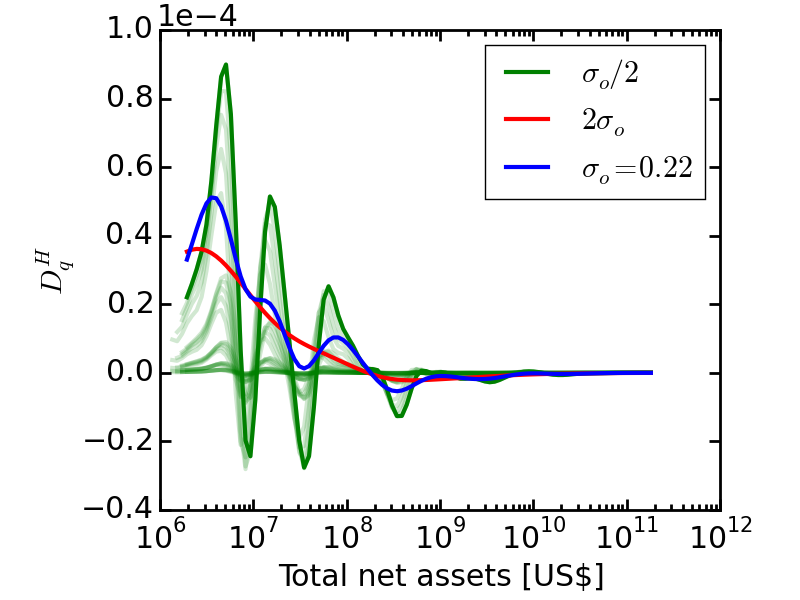}
\caption{The left figure shows the Gaussian KDE of the PDF of the  logarithms of the ETF sizes for three different values of the bandwidth. The blue line is for the optimal bandwidth ($\sigma_{o}=0.22$) determined using cross validation. The green and red line correspond with a bandwidth of $\sigma_{o}/2$ and $2 \sigma_{o}$. The right figure shows the generalized derivative $D^{H=0.5}_{q=0.65}$ of the curves of the left figure. For $\sigma_{o}/2$ we also show $D^H_q$ for different combinations of the values $(0.5 \le H \le 0.9, 0.65 \le q \le 0.95)$.} 
\label{fig:KDEofLogData}
\end{figure}

Working with the PDF of ETF sizes has the advantage compared with the CCDF of being a local measure of the distribution, hence less prone to the influence of contamination by systematic biases.
However, the PDF is more noisy and harder to estimate with limited data. A standard and robust estimation method consists in constructing its kernel density estimator, which is a kind of smoothed histogram. The Gaussian KDE of the PDF of the logarithms $\ln S$ of ETF sizes  is defined as
\begin{equation}
\widehat{f}_{\sigma} \left( \ln S \right) = \frac{1}{N} \sum^{N}_{i=1} \mathcal{N} \left( \ln S - \ln S_i, \sigma ^2 \right) \; ,
\end{equation}
with $\mathcal{N}(0,\sigma ^2)$ a zero-mean Gaussian distribution with variance $\sigma^2$,  and the sum is over the $N=479$ data points $\ln S_i$. Further, in the context of KDE one refers to  $\sigma$ as the bandwidth.  Figure~\ref{fig:KDEofLogData} shows the KDE of the PDF of ETF sizes for different bandwidths $\sigma$. The optimal bandwidth $\sigma_{o}=0.22$ is determined with cross validation. Recall that, in cross validation, the model is first fit to part of the data, after which a quantitative metric is computed to determine how well this model fits the remaining data. Obviously, there are strong indications for oscillatory behavior emerging from the KDE analysis of the PDF of ETF sizes. We use the generalized derivative of this function in order to gain a better insight into this oscillatory behaviour.  
 

\begin{figure}[htb]
\centering
\includegraphics[scale=0.5, trim={0 0cm 0 0cm},clip]{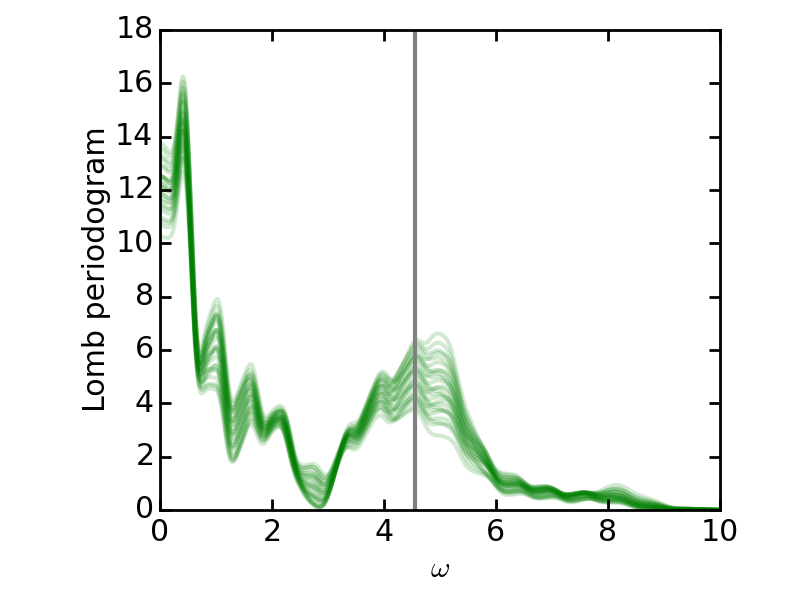}
\caption{Lomb periodogram of the generalized $(H,q)$-derivative of the Gaussian KDE of the PDF of the decimal logs of the ETF sizes with $\sigma_{o}/2$ and  different values of the combination $(0.5 \le H \le 0.9, 0.65 \le q \le 0.95)$. The black vertical line is at the center value of $\omega = 4.6 \pm 0.6$, which corresponds to the scaling ratio for the ETF sizes $S$ of $p = \exp (2\pi / \omega ) = 3.9 \pm 0.4$.}
\label{fig:LombPeriodogram}
\end{figure}

The generalized $(H,q)$-derivative of a function $f(x)$ is defined as \cite{zhou2002generalized,zhou2003nonparametric}
\begin{equation}
D^{H}_{q}f(x) \equiv \frac{f(x)-f(q x)}{ \left[ (1-q) x \right]^{H} }~,
\end{equation}
and provides a robust metric of the trend or slope of a function. This is particularly useful to detect features in a noisy function, such as the PDF of ETF sizes studied here. Figure~\ref{fig:KDEofLogData} includes the $D^{H}_{q}$ of the KDE of the PDF of ETF sizes for three bandwidths.  As recommended in Refs.~\cite{zhou2002generalized,zhou2003nonparametric}, we have scanned $H$ from $0.5$ to $0.9$ in steps of $0.08$, and $q$ from $0.65$ to $0.95$ in steps of $0.06$ and found that the results are robust. Accordingly, the displayed  $D^{H=0.5}_{q=0.65}f(\ln S)$ results can be considered representative.  One can observe three to four well formed oscillations in the logarithm of the ETF sizes $S$, quite similarly to the observations of the cumulative distribution approach. 


In order to extract the strongest contributing frequencies, we have computed the Lomb periodogram of the generalized $(H,q)$-derivative of the Gaussian KDE of the PDF of the logarithms of the ETF sizes. We choose the kernel estimation with $\sigma_{o}/2$ as it is representative of the other estimators but exhibits the largest oscillatory amplitudes.  The resulting periodograms are shown in Fig.~\ref{fig:LombPeriodogram}.  First, the peaks at low angular log-frequencies $\omega < 1.5$ represent oscillations with a wavelength of about the size of the entire range of values and hence can be ignored as explained above \cite{Huangetalarti00}. There is only one noticeable peak at a value of $\omega =4.6 \pm 0.6$ that can be put in correspondence with the second harmonic $\omega_2 = 5.2 \pm 0.2$ previously reported. This angular log-frequency corresponds to a scaling ratio of $p_2 = \exp (2\pi / \omega_2 ) = 3.9 \pm 0.4$.  Note that, when averaging the Lomb periodogram over the scanned $H$ and $q$, the same estimate $\omega =4.6 \pm 0.6$ for the unique significant peak is obtained, providing
evidence that it has a real existence. There is no significant peak at $\omega_1$, likely as a result of the high-frequency noise associated with the construction of the PDF. Note that the general available theory of log-periodic functions indicates that different harmonics can have very different amplitudes that depend on subtle properties of the problem \cite{GluzmanSornette02}. In other words, one should not be surprised that the different harmonics of log-periodicity express themselves with different amplitudes in distinct signals.

\subsection{Mechanisms of discrete scale invariance in the PDF of ETF sizes}
\label{subsec:mechanisms}

Sections~\ref{wrnhjki} and \ref{qettb} have presented statistically significant evidence of 
the existence of a discrete hierarchical structure in the distribution of ETF sizes, 
with preferred scaling ratios approximately equal to $p_1 \approx 12$ and $p_2 = \sqrt{p_1} \approx 3.5$. Reference~\cite{SornetteDSI98} provides a review of the many mechanisms that can produce
such a discrete hierarchy. While we cannot offer a definite mechanism and test for its relevance,
the most likely candidates are the Kesten process \cite{sornette1998linear,jogi1998fine}
and aggregation/fragmentation dynamics \cite{ouillon1996imaginary}.
The Kesten process can be used to describe the growth of portfolio as a result of the joint addition 
of new deposits and of stochastic proportional growth.  Log-periodic PDFs emerge quite robustly if the multiplicative stochastic factors are not too broadly distributed. The aggregation/fragmentation dynamics could be also a limiting process for the formation of ETF portfolios, for which it can be shown that discrete scale invariance may emerge for quite general aggregation/fragmentation kernels.

Another explanation for the occurrence of distributions with a power law decorated by log-periodic oscillations finds its origin in the formalism of nonextensive statistical mechanics \cite{ISI:A1988Q240100029}. At any instant of time, the universe of ETFs behaves as an open system that seeks to find equilibrium with the whole of the capital market that acts as a reservoir. In the context of equilibrium statistical physics, this equilibrium process \cite{Dragulescu2000} gives rise to the well-known Boltzmann-Gibbs distribution of the sizes of the ETFs 
\begin{equation}
P^{BG} \left( 0 \le S \le \infty \right) = \frac {1} {T_0} \exp - \frac {S} {T_0} \; ,
\label{eq:SasBoltzmann}
\end{equation}
where the temperature $T_0$ acts as a typical scale parameter for the size of the ETFs. In this picture, all ETFs (independent of size) are subject to a similar stochastically driven capital exchange with the reservoir of the complete market. In other words, the universe of ETFs is embedded in the reservoir of investment products and the exchange between any element in the ETF universe and the reservoir can be parametrized by a single scale parameter $T_0$. The larger $T_0$ the larger the average size of the ETFs. Obviously, Eq.~(\ref{eq:SasBoltzmann}) does not give rise to fat tails in the distribution of the ETF sizes and cannot be considered realistic given the observations of Section~\ref{yjuthgdqa}.
A generalization, however, proceeds as follows. The Boltzmann-Gibbs exponential distribution (\ref{eq:SasBoltzmann}) is a solution to the following equation
\begin{equation}
\frac { d P^{BG} (S)} {dS} = - \frac {1}{T_0} P^{BG} (S) \; .
\label{eq:diffBG}
\end{equation}
In nonextensive statistical mechanics, this equation is extended by adding a nonextensivity  parameter $n$
\begin{equation}
\frac { d P (S)} {dS} = 
- \frac {1}{T(S)} P (S) = 
- \frac {1}{T_0 + \frac {S} {n}} P (S) \; .
\label{eq:Tsallis}
\end{equation}
A particular solution to this equation is known as the Tsallis distribution
\begin{equation}
P(S) = \frac {n-1}{n T_0} \left( 1 + \frac {S} {n T_0} \right) ^ {-n} \; .
\label{eq:TsallisDis}
\end{equation}
The Tsallis distribution nicely interpolates between the Boltzmann-Gibbs exponential for ETF sizes $S$ smaller than the scale parameter $T_0$ and a power law tail for $S \gg T_0$.   
The nonextensivity parameter $n$ makes the temperature --in the current context a proxy for the typical amount of capital exchange of an ETF with the reservoir of investment products-- dependent on the actual ETF size $S$. Loosely speaking, the parameter $\frac{1}{n}$ can be interpreted as a measure for the degree of preferential attachment \cite{2005EL7070S}, or the extent to  which the rich ETFs get richer. In the current context, the $n$ accounts for the fact that there is an increased linear tendency of an ETF to accrue money from the reservoir of investment products as it increases in size $S$. The quantity $n$ determines the asymptotic behaviour of the distribution $P(S)$. The normalization condition $\int _{0} ^{\infty} P(S) d S$ of the distribution (\ref{eq:TsallisDis}) requires that $n>1$.  The Tsallis distribution, for example, provides an excellent fit to transverse momentum distributions in high-energy collisions with values of $n$ of the order 6-8 \cite{Cleymans:2015lxa}. Obviously, the limit $n \rightarrow \infty$ corresponds to a vanishing preferential attachment effect. The smaller $n$  the larger the difference between the temperature associated with the small and the large ETFs. Small ETFs, that are defined as those with a current size smaller than the scale $T_0$ experience a temperature $T  \approx T_0$ in their interaction with the capital market.  Large ETFs, defined as ETFs larger than the scale $T_0$, experience a temperature $T(S)$ that scales linearly with their size  $T (S) = T_0 + \frac {S} {n}$.

Building on the connection between preferential attachment growth and nonextensive statistical mechanics \cite{2005EL7070S} and following the derivations of  Ref.~\cite{ISI:000348403600019} we now explain that the differential equation of the type (\ref{eq:Tsallis}) can give rise to distributions $P(S)$ that have a power law tail decorated with log-periodic oscillations if one adds an evolutionary aspect to the system. In finite difference form, the Eq.~(\ref{eq:Tsallis}) can be written as
\begin{equation}
P(S+ \delta S) = \frac {-n \delta S + n T_0 +S}{n T_0 +S} P(S) = \frac {-\delta S + T(S)}{T(S)} P(S) \; ,
\label{eq:deriv1}
\end{equation}
where $\delta S$ can be interpreted as a single-step small increment of the ETF size $S$. We now seek to find the solutions to the evolution equation (\ref{eq:deriv1}) for a specific choice for the increment $\delta S$. As the changes $\delta S$ can be anticipated to be proportional to the fluctuating temperature $T(S)$ one can introduce an additional scale parameter $\gamma$
\begin{equation}
\delta S \equiv \gamma n T(S) = \gamma n \left( T_0 + \frac {S} {n} \right) = \gamma n T_0 \left( 1 + \frac {S} {n T_0} \right) \; ,
\label{eq:deriv2}
\end{equation}
where $\gamma$ can be made arbitrary small by imposing the condition $\gamma \ll \frac {1} {n}$ and recalling that $n>1$. After inserting the expression (\ref{eq:deriv2}) into (\ref{eq:deriv1}) one finds that 
\begin{equation}
P \left(  S (1+ \gamma) + \gamma n T_0 \right) =  (1 - \gamma n ) P(S) \; .
\label{eq:deriv4}
\end{equation}
In the asymptotic regime $S \gg T_0$, one finds
\begin{equation}
P \left(  S (1+ \gamma) \right) \approx  (1 - \gamma n ) P(S) \; \hspace{0.03\textwidth} (S \gg T_0) ,
\label{eq:deriv5}
\end{equation}
an expression that for finite values of $\gamma$ is directly recognized as the usual condition $P (\lambda S) = \mu P(S)$ for scale invariance of the function $P(S)$. It is well known \cite{SornetteDSI98,ISI:000348403600019} that the most general solution for the asymptotic part of the distribution is a linear combination of power laws with complex exponents $\alpha _k (\gamma,n)$
\begin{equation}
P(S) \approx \sum _{k \in \mathbb{N}} w_k S ^{-\alpha _k (\gamma,n)} \hspace{0.03\textwidth} (S \gg T_0) ,
\end{equation}
with,
\begin{equation}
\alpha _{k \in \mathbb{N} } (\gamma,n) = - \frac {\ln \left( 1 - n \gamma \right) } {\ln \left( 1 +  \gamma \right)}
+ \frac {2 \pi i k} {\ln \left( 1 +  \gamma \right)} \; .
\end{equation}
As is usually done, we retain only the terms in $w_0$ and $w_1$ and the real part of the function, to obtain
\begin{equation}
P (S)    \sim   S^{-n - \frac {n}{2} (n+1) \gamma + \mathcal{O} \left( \gamma ^ 2 \right)}  
\left[w_0 + w_1 \cos \left( \frac {2 \pi} {\ln (1+ \gamma)} \ln S \right)
\right]  \hspace{0.03\textwidth} (S \gg T_0) \;.
\label{eq:thefinalg}
\end{equation}
For large values of the ETF size $S$ ($S \gg T_0$), the distribution $P (S)$ behaves as a power law decorated with a log-periodic oscillation of the type  $\Delta F(S)$ defined in  Eq.~(\ref{rnwrgbvqd}). This is compatible with the qualitative findings for the tail parts of the empirical distribution of ETF sizes (see Figs.~\ref{fig:CCDFandLognormal} and \ref{fig:KDEofLogData}). We stress that the log-periodic oscillation in the above distribution $P(S)$ is determined by the finite parameter $\gamma$ that is connected with the time evolution of the system in accordance with multiplicative size increments $\delta S$ that obey the relation (\ref{eq:deriv2}). 
For infinitesimally small increments -- that correspond with $\gamma \rightarrow 0$ -- one has that
\begin{equation}
\lim _{\gamma \to 0}e ^ {- \alpha _ k (\gamma,n)}  = e ^ {- \alpha _0} \;  \; \; \; (\forall k) \; , 
\end{equation}
and the asymptotic distribution of~(\ref{eq:thefinalg}) reduces to the tail  $S^{-n}$ of the Tsallis distribution (\ref{eq:TsallisDis}).   

As a matter of fact, the proposed asymptotic solution (\ref{eq:thefinalg}) of the evolution equation, provides one with a prediction for the angular frequency of the oscillations in $\ln S$ after one time step
\begin{equation}
\omega _1 = \frac {2 \pi} {\ln (1+ \gamma)} \; .
\end{equation}
The measured distribution of ETF sizes is the result of many multiplicative evolution steps of the type (\ref{eq:deriv1}) each with its finite characteristic scale parameter $\gamma_t$.   
The size of the ETF at the time instances $t$ and $t - \Delta t$ are connected by an expression of the type (\ref{eq:deriv2})
\begin{equation}
\delta S _t = S_t - S_{t -\Delta t} = \gamma _t n \left( T_0 + \frac {S_{t-\Delta t}} {n} \right) \; .
\label{eq:multi1}
\end{equation}
For the sake of simplicity, let us assume that there are $\kappa$ time steps $\Delta t$ and that all $\gamma _t$ are equal: $\gamma _t = \gamma, \forall t$.

Proceeding in a fashion analogous to the above derivations and detailed in Ref.~\cite{ISI:000348403600019}, one finds after $\kappa$ time steps an asymptotic distribution $P(S)$  that is similar to the result of (\ref{eq:thefinalg}) apart from the following substitution in the angular frequency $\omega_{\kappa}$ of the $\cos \left( \omega \ln S \right) $ term
\begin{equation}
\omega _1 = \frac {2 \pi} {\ln (1+ \gamma)} \Longrightarrow
\omega _{\kappa} = \frac {2 \pi} {\kappa \ln (1+ \gamma)} \; .
\end{equation}
This means that the angular frequency of the oscillations in $\ln S$ decrease with the finite number of time steps $\kappa$ as $\frac {1} {\kappa}$. As a consistency check and referring to the observed oscillations in the size distributions of the ETFs in  Fig.~\ref{fig:LombPeriodogram}: for $\omega =4.6$ one finds $\gamma = 0.014$ for $\kappa = 100$ and $\gamma = 0.0014$ for $\kappa = 1000$.   
\begin{figure}
\centering
\includegraphics[scale=0.5, trim={0 0 0 0},clip]{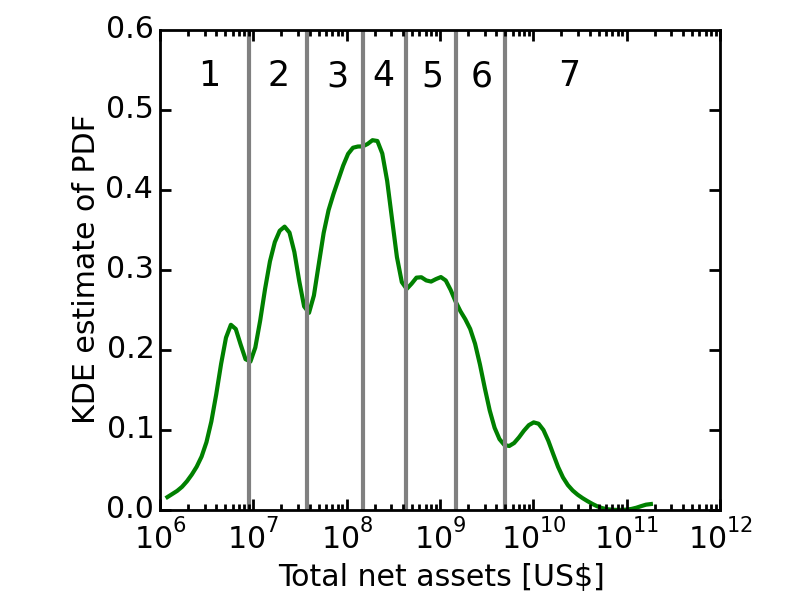}
\caption{Partitioning of the distribution of ETF sizes obtained with the Gaussian KDE of the PDF of the  logarithms of the ETF sizes with $\sigma_{o}/2$, by identifying  the minima and maxima that are separated by a factor close to $p_2 \approx 3.5$. The seven size layers are bracketed by the vertical lines.}
\label{fig:KDEofLogData_scales}
\end{figure}

\section{Analysis of the economic significance of the hierarchy of ETF sizes}
\label{sec:significance}

We now explore whether the discrete hierarchical structure in the distribution of ETF sizes could be associated with  economic properties of the ETF portfolios. Studying the return-risk properties of the hierarchy of ETF sizes is tantamount to investigating the generalisation of the size factor (also often referred to as SMB for ``Small [market capitalization] Minus Big [market capitalization]'') of the three-factor Fama-French model \cite{FamaandFrench1993}. Recall that the introduction of the SMB factor was motivated by the observation that small capitalisation stocks have tended to do better than the market as a whole. 
The observation that the size effect is rather weak, especially in the last decade, has pushed Fama and French to extend their three-factor model to a five-factor model \cite{FamaandFrench2015}. Therefore, we expect to find only weak signatures of the size hierarchy. Nevertheless, we propose that it is worthwhile to investigate a generalisation of the dichotomy between small and big ETF sizes, by using the discrete hierarchy discovered above.

In order to construct the size layers partitioning the ETF universe, we use a specific geometric partitioning of the ETF universe based on the discovered scaling ratio $p_2 =  \exp (2\pi / \omega_2 ) \approx \sqrt{p_1} \approx 3.5$, because it is present both in the analysis of the CCDF (Section~\ref{wrnhjki}) and of the PDF (Section~\ref{qettb}) of ETF sizes. Moreover, it amounts to the simplest substructure to the dominant scaling ratio $p_1 \approx 12$ identified in Section~\ref{wrnhjki}. We partition the distribution of ETF sizes obtained with the Gaussian KDE of the PDF of the logarithms of the ETF sizes with $\sigma_{o}/2$, by identifying  the minima and maxima that are separated by a factor close to $p_2 \approx 3.5$. The obtained set of seven size layers are represented in Fig.~\ref{fig:KDEofLogData_scales}. Table~\ref{tab:bandProperties} reports a number of properties for each size layer $i$, including the number of ETFs, the average number of holdings per ETF, the upper bound size ($ub_{i}$) and the ratio $ub_{i} / ub_{i-1}$. One can observe that the mean value of this ratio is $3.6$, which is close to the scaling ratio of $p_2 = 3.5 \pm 0.2$, as expected. Note that the most probable ETF size of approximately 130 million US\$ determined in Section~\ref{yjuthgdqa} falls close to the boundary between the third and fourth size layer. In contrast, the mean ETF size of 1.6 billion US\$ is close to the boundary between the fifth and sixth size layer.

\subsection{Intra-layer and inter-layer similarity of stock holdings across ETF size layers}
\label{subsec:intra_and_inter_similarity}

\begin{figure}
\centering
\includegraphics[scale=0.5, trim={0 0 0 0},clip]{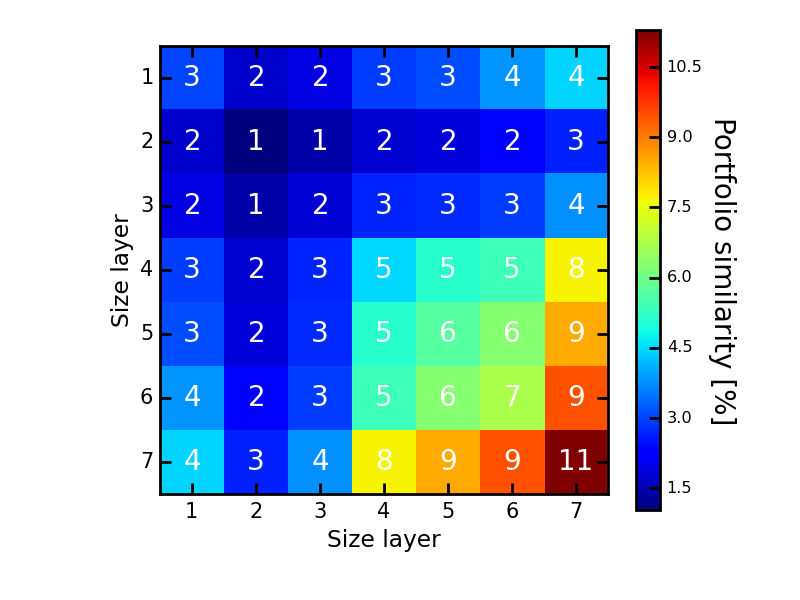}
\caption{Matrix of the average intra-layer and inter-layers similarities of ETFs across the seven size layers defined from Fig.~\ref{fig:KDEofLogData_scales} and detailed in Table~\ref{tab:bandProperties}. An entry SIM$(i,j)$ of this matrix, as indicated by the white number and the color scale, is the average similarity expressed in percentages between the portfolios of all ETFs in size layer $i$ with all the ETFs in size layer $j$. The similarity between two ETFs is defined by Eq.~(\ref{eq:portfolio_similarity}). 
}
\label{fig:overlap_NoPossibleOverlaps_cos}
\end{figure}

To investigate whether there is a connection between the different size layers and the portfolio composition of the ETFs, we compare the portfolio similarity of the different scales. The portfolio similarity $sim_{ee^{\prime}}$ of  ETFs $e$ and $e^{\prime}$ is defined as
\begin{equation}
sim_{ee'} = sim_{e'e} = \frac{\sum_{i \in |h_e \cap h_{e'}|} w_{ei} w_{e'i}}
{	\sqrt[]{\sum_{k \in h_e} w_{ek}^{2} } \quad
	\sqrt[]{\sum_{l \in h_{e'}} w_{e'l}^{2} }},
\label{eq:portfolio_similarity}
\end{equation}
where $h_e$ and $h_{e'}$ are all the holdings of ETFs $e$ and $e'$, and $w_{ei}$ is the portfolio weight of holding $i$ in ETF $e$.

Figure~\ref{fig:overlap_NoPossibleOverlaps_cos} represents the matrix of average intra-layer and inter-layer similarities of ETFs, as defined by (\ref{eq:portfolio_similarity}) across the seven size layers. Specifically, an entry SIM$(i,j)$ of this matrix is the  average similarity $sim_{ee'}$ between the portfolios of all ETFs $e$ in size layer $i$ with all the ETFs $e'$ in size layer $j$.  Firstly, one sees that size layers consisting of larger ETFs are more self-similar. As the size layer number $i$ increases, there is less diversity in the number of holdings used to construct the portfolios of the corresponding ETFs. The SIM$_{I} \equiv$SIM$(I,I)$ column of Table~\ref{tab:bandProperties} reports these intra-size layer similarities. To sum up, two size layers of large ETFs are more similar than two size layers of smaller ETFs or than a size layer of small ETFs and a size layer of large ETFs. The first smallest size layer $1$ breaks this regularity, which is kind of an oddity that can perhaps be associated with the very small sizes of these ETFs. The column SIM$_{M}$ of Table~\ref{tab:bandProperties} reports the average overlap of the ETFs in a given size layer with the market portfolio. To construct the market portfolio, we consider all the stocks held by ETFs. The weight of a stock in this portfolio is simply its market capitalization divided by the total market capitalization of all the stocks combined. Not surprisingly, one can observe that the larger size layers exhibit a stronger similarity to the market portfolio.

\begin{table}[htb]
\begin{center}
\caption{For each identified ETF size layer $i=1,2,\ldots, 7$, this table reports the corresponding interval of covered market capitalisations with the upper ($ub_{i}$) and lower bound ($lb_{i}$). For example, size layer 2 contains ETFs with capitalisations between 9$\times 10^6$~US\$ and 38$\times 10^6$~US\$.  Further, for each size layer we provide the number of ETFs, the average number $\bar{N}_{h}$ of holdings per ETF, the ratio $ub_{i} / ub_{i-1}$, the average overlap similarity over all ETF pairs in a band (SIM${I}$) and the average overlap with the market portfolio (SIM$_{M}$). 
}
\label{tab:bandProperties}
\begin{tabular}{ c  c  c  c  c  c }
\hline \hline
  Size layer & \#ETFs & $\bar{N}_{h}$ & $ub_{i} / ub_{i-1}$ & SIM$_{I}$ $(\%)$ & SIM$_{M}$ $(\%)$ \\
  $\left] lb_i, ub_i \right]$ $\bigl( 10^6$US\$ $\bigr)$ & & & & \\   
  \hline
  1: $\left] 0, 9 \right] $ & 48  & 172  & & 3.03 & 13.04\\
  2: $\left] 9, 38 \right]$ & 88  & 150  & 4.2 & 1.04 & 7.78\\
  3: $\left] 38, 150 \right]$ & 109 & 183 & 3.9 & 1.82 & 8.54\\
  4: $\left] 150, 430 \right]$ & 84  & 229 & 2.9 & 4.5 & 14.37\\
  5: $\left] 430, 1500 \right]$ & 77  & 258 & 3.4 & 5.72 & 17.58\\
  6: $\left] 1500, 5000 \right]$ & 43  & 281 & 3.3 & 6.74 & 19.44\\
  7: $\left] 5000, \infty \right]$ & 30  & 288 & & 11.29 & 26.13\\
  \hline \hline
\end{tabular}
\end{center}
\end{table}

\begin{figure}[htb]
\hspace{0.20\textwidth} \Large{$\mathcal{M}^{bin}_{bh}$} 
\hspace{0.45\textwidth} \Large{$\mathcal{M}^{frac}_{bh}$} \\
\includegraphics[scale=0.40, trim={0 0 0 0},clip]{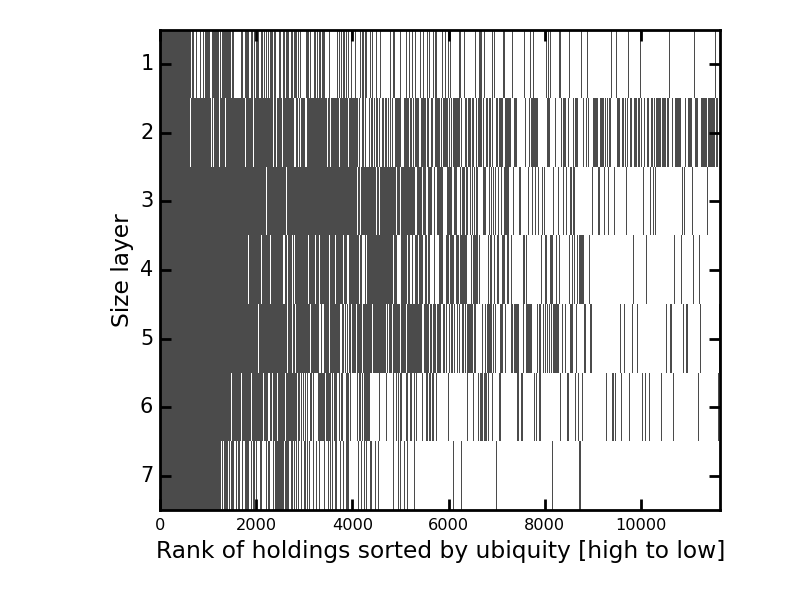} 
\includegraphics[scale=0.40, trim={0 0 0 0},clip]{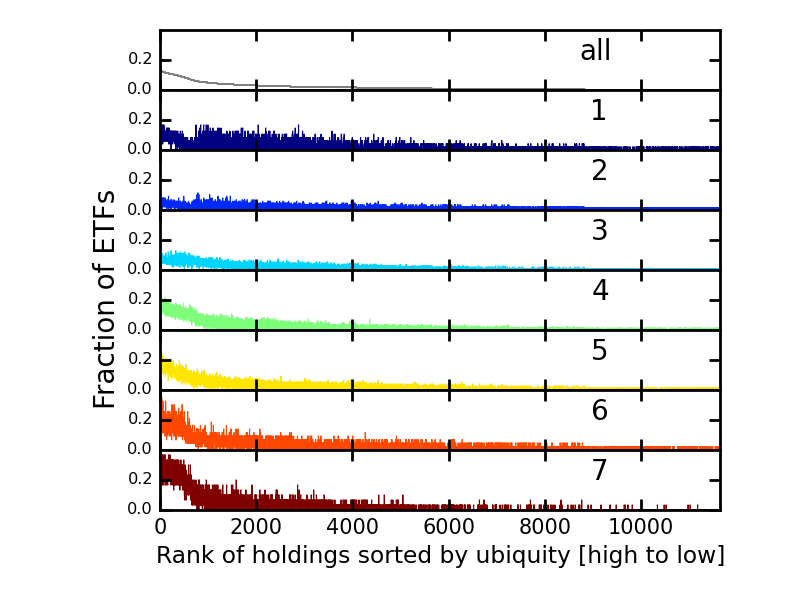}
\caption{The matrices $\mathcal{M}^{bin}_{bh}$ (left) and $\mathcal{M}^{frac}_{bh}$ (right) as defined in the text. The upper panel  of  $\mathcal{M}^{frac}_{bh}$ shows the fraction of the equity ETFs in which a certain holding occurs. The 11,643 holdings are sorted according to their ubiquity with rank 1 corresponding to the most ubiquitous stock.}
\label{fig:adjacency_matrix}
\end{figure}

The greater intra-layer and inter-layer similarity of ETFs of large sizes is not surprising, as a large amount of capital to invest needs to find a large number of potential firms with not too large weights in order to limit market impact. As the universe of available stock investment is finite, and the set of attractive stocks is even more limited at any given time, it can be expected that the large ETFs exhibit significant overlaps in their holdings. For instance, who would not hold the largest firms such at Apple in their portfolio?

To quantify further this similarity in the holdings of the large ETFs, we define the two adjacency matrices $\mathcal{M}^{bin}_{bh}$ and $\mathcal{M}^{frac}_{bh}$ with dimensions $(N_b \times N_h)$. Here, $N_b=7$ is the number of size layers and $N_h=11,643$ is the total number of distinct holdings over the 479 equity ETFs considered. The fact that one ETF in the size layer $b$ has a position in stock $h$ is encoded by $\mathcal{M}^{bin}_{bh} = 1$, otherwise $\mathcal{M}^{bin}_{bh} = 0$. The second matrix is defined such that the element $\mathcal{M}^{frac}_{bh}$ is equal to the fraction of ETFs in the size layer $b$ that have a position in the holding $h$. The adjacency matrices $\mathcal{M}^{bin}_{bh}$ and $\mathcal{M}^{frac}_{bh}$ are shown in Fig.~\ref{fig:adjacency_matrix}. The holdings are sorted from highest to lowest ubiquity in the 479 equity ETFs considered in our analysis. The adjacency matrices of Fig.~\ref{fig:adjacency_matrix} allow us to draw several conclusions. First, larger ETFs tend to use a smaller set of stocks to invest in. Second,  larger ETFs tend to select increasingly from the same ubiquitous stocks. The ETFs in size layer 2, on the other hand, nicely sample from the entire space of holdings. Third, the first size layer $1$ is an exception to this stylized picture and appears to be a smaller version of size layers 5 and 6.

\subsection{Relationship between stock holding ubiquity and capitalisation within ETF size layers}
\label{subsec:ubiquity_vs_cap}

\begin{figure}
\begin{center}
   \noindent
\includegraphics[width=1.\textwidth,trim=0cm 0cm 0cm 0cm, clip=true]{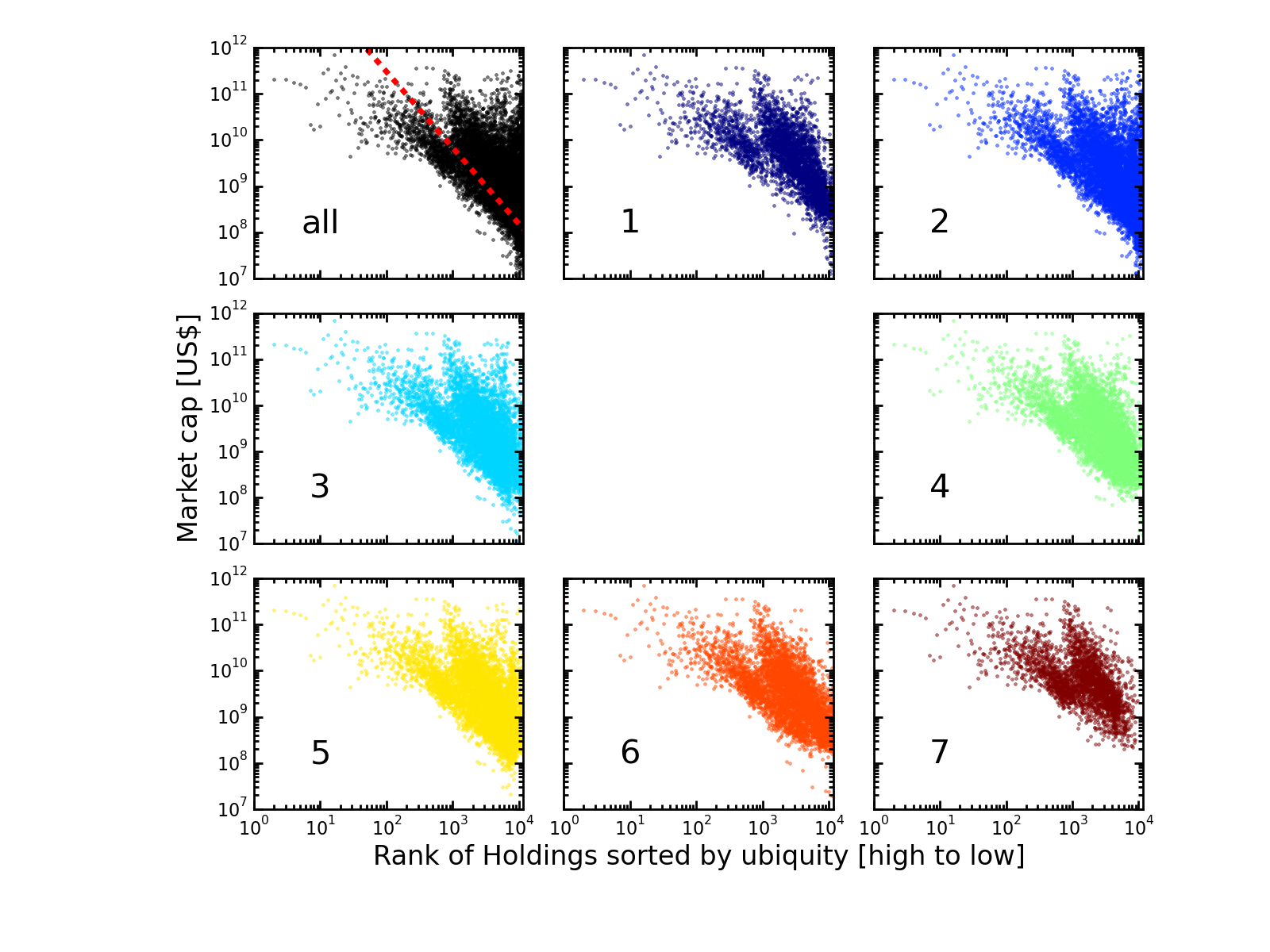}
\end{center}
\caption{For all the 479 ETFs and for each size layer separately, the market capitalization at the end of 2014 for all the holdings considered in Fig.~\ref{fig:adjacency_matrix}. The holdings are sorted from highest to lowest ubiquity and the results are plotted on log-log axes (base 10). The dashed red line in the upper left panel indicates the separation between the two clusters mentioned in the text.
}
\label{fig:sortedHoldingsMarketCap_log}
\end{figure}

\begin{figure}
\centering
\includegraphics[scale=0.5, trim={0 0 0 0},clip]{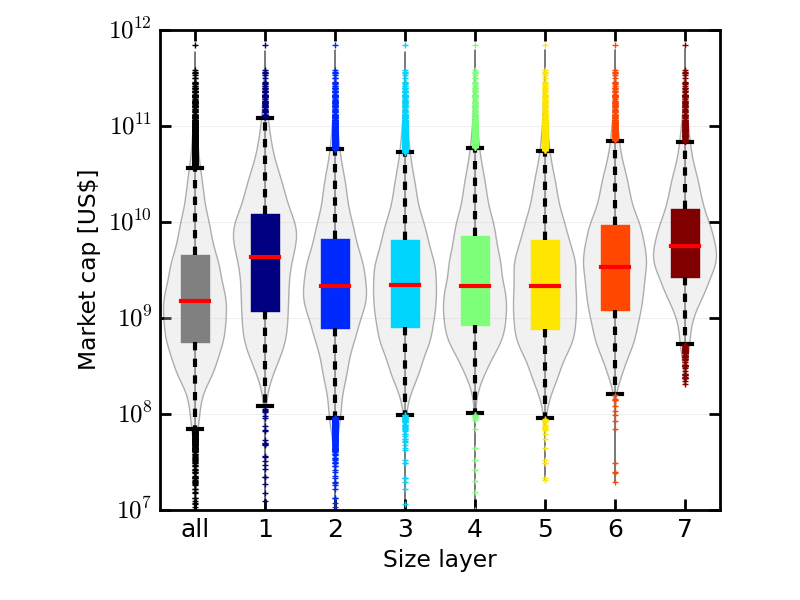}
\caption{For the entire sample of 479 equity ETFs and for each size layer separately, this figure shows the boxplots of the market capitalisations of all the stocks held by the ETFs. We note that the boxplot covering all the holdings is not a mere aggregation of all the size layers. In addition to presenting the boxplots, we show also the so-called violin plots that give the full distribution in thin lines along each vertical axis.}
\label{fig:MarketCap_boxplot}
\end{figure}

As already mentioned, an obvious explanation for why larger ETFs tend to be more similar is that they need to hold stocks with a larger market capitalisation. Holding too many smaller stocks might prove too costly and not sufficient to absorb the capital in need of investment opportunities. The concentration and similarity of large ETFs may, in large part, just reflect the reduction in available large stocks to invest in. To investigate this hypothesis, we first study the relation between the ubiquity of a stock and its market capitalisation. The results are shown in 
Fig.~\ref{fig:sortedHoldingsMarketCap_log} for all holdings and per size layer respectively. 
The figure exemplifies that, up to the 500 largest firms (up to rank 500), the sizes of the corresponding firms are drawn from approximately the same distribution with a minimum size of about 5$\times$10$^9$ US\$. In contrast, below rank $\approx$500, one can observe a simple power law relationship relating the size of the smallest admissible firms with respect to their abundance in the universe of ETFs.  The corresponding exponent of the power law is $\alpha = 1.7028 \pm 0.0002$ where the power law is $f(x) \propto 1/x^{\alpha+1} $.

As expected, smaller stocks tend to be less ubiquitous in large ETFs. This effect is most obvious in the panel of size layer seven. On the contrary, we do not see that small ETFs only hold small cap stocks. Surprisingly, Fig.~\ref{fig:sortedHoldingsMarketCap_log} uncovers some additional structure. There appear to be two clusters in the considered ``market cap''-``holding ubiquity'' matrix,  and to guide the eye we have drawn a separation line in the panel including all stocks.
 
\begin{figure}[htb]
\centering
\includegraphics[scale=0.4, trim={0 0 0 0},clip]{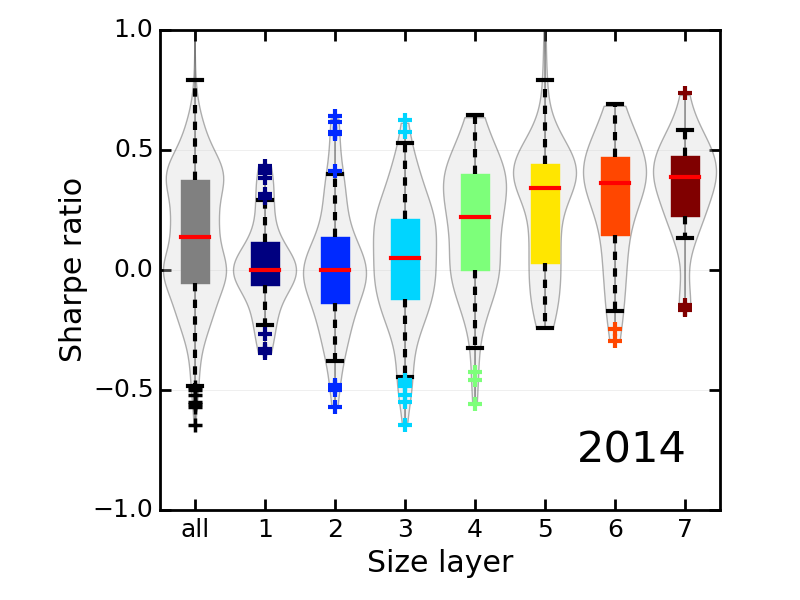}
\includegraphics[scale=0.4, trim={0 0 0 0},clip]{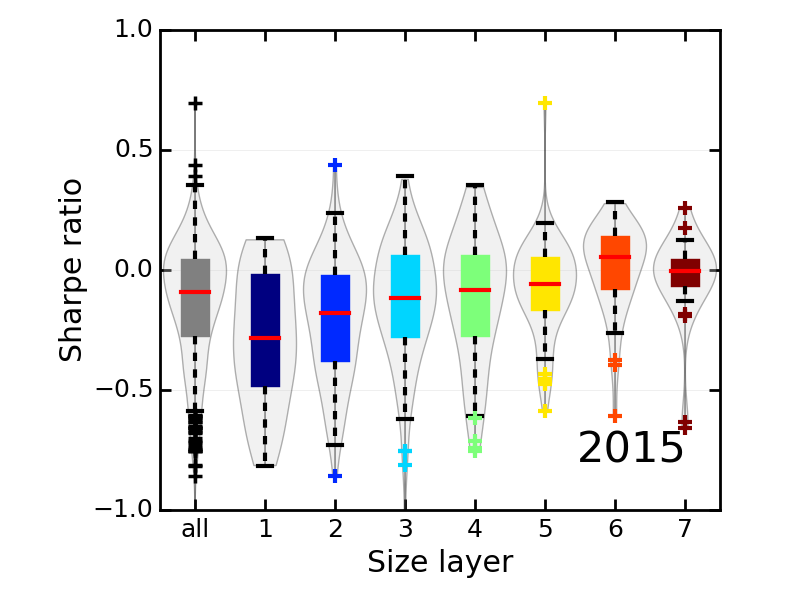}
\caption{Performance for the entire sample of ETFs and the different size layers in the years 2014 (left) and  2015 (right). We present the performance measure as boxplots complemented by violin plots giving the full distribution in thin lines along each vertical axis.
}
\label{fig:performance}
\end{figure}

We have also investigated the distribution of the market capitalisation of all the stocks that appear in the discerned size layers. Figure~\ref{fig:MarketCap_boxplot} shows the market capitalisations of all the stocks that are held by the ETFs in the considered subset, be it the entire sample of ETFs or a particular size layer. The boxplot for all the holdings over all ETFs is not a mere aggregation of the boxplots of all size layers, as stocks can appear only once. For size layers 6 and 7, but remarkably also for size layer 1,  the distribution is clearly shifted to stocks with a larger market capitalisation.  
The violin plots of the full PDFs for the whole ETF market and each size layer reveal that the distributions have more structure than being uni-modal and symmetric. They exhibit significant skewness as well as bi-modality as is the case for size layers 1 and 5.

\subsection{Investment performance across the seven ETF size layers}
\label{subsec:performance_over_layers}

Figure~\ref{fig:performance} presents a measure of performance of the ETFs in each size layer and in total for 2014 and 2015. Unfortunately, we do not have access to comparable data on ETFs for other calendar years, which prevents us from performing factor regressions as in 
\cite{FamaandFrench1993,FamaandFrench2015}. As metric of performance, we use the Sharpe ratio, defined as the annualised mean return divided by annualised volatility (standard deviation of the returns). We take a reference risk free interest rate equal to $0$. In addition to presenting the standard boxplots, we show also the violin plots that give the full distribution in thin lines
along each vertical axis.
 
In 2014, all median Sharpe ratios were positive, showing that the majority of ETFs generated positive returns. In addition, a significant inverse-size effect can be observed: the upper size layers 5-7 significantly 
over-perform the lowest size layers 1-3, with size layer 4 representing an intermediary case. In contrast, 2015 has been a difficult year for ETFs as well as for hedge-funds in general. Except for size layer 6, all other size layers have negative median Sharpe ratios. One can also observe an inverse size effect, in the sense that the upper size layers also perform better than the lower ones, but the difference is less pronounced than in 2014. The higher performance of large ETFs is reminiscent of the increasing returns to wealth inequality found, for example, for university endowments \cite{Piketty2014} and for households' portfolios \cite{Campanale07}: larger endowments provide much larger returns as a result of better economies of scales, and of access to more investment opportunities and to more skilled managers. Similarly, larger households' wealth enable access to more diversified portfolios.

The violin plots of the full PDFs for the whole ETF market and each size layer in Fig.~\ref{fig:performance} reveal that the distributions have more structure than being uni-modal and symmetric.  The discerned bi-modality of the PDF of Sharpe ratios for the whole market can be attributed to the distinct performance of two classes of ETFs, the less performing one represented mostly in size layers 2-4
and the more performing one populating the size layers 5-7.

\section{Conclusion}
\label{sec:conclusion}

We provided a novel detailed analysis of the size distribution of a universe of almost 500 equity ETFs and discovered a discrete hierarchy of sizes, which imprints a log-periodic structure on the probability distribution of ETF sizes that dominates the details of the asymptotic tail. We used the found discrete hierarchy to propose a classification of the whole universe of ETFs into seven size layers. Introducing a similarity metric, we found that the largest ETFs exhibit both stronger intra-layer similarity and stronger inter-layer similarity compared with the smaller ETFs. We have found strong indications that this  reflects the obligation for large ETFs to  spread their capitalisation on a relatively more reduced set of large stocks. This lack of diversification in the classes of large ETFs seems to reinforce
the concentration of stock capitalisation known as Zipf's law. This concentration together with the similarity of holdings suggests potential vulnerability to systemic risks. We also provided comparative performance
across the seven ETF size layers and found an inverse size effect, namely large
ETFs perform significantly better than the small ones. 

\section*{Acknowledgments}
This work was supported by the Research Foundation Flanders (FWO-Flanders). B.~Vandermarliere was supported as an `FWO-aspirant'.  

\newpage 
\section*{Bibliography}

\bibliographystyle{unsrt}
\bibliography{bibliography}

\begin{thebibliography}{10}

\bibitem{ETFRisk1}
I~Diaz-Rainey and G~Ibikunle.
\newblock {A Taxonomy of the ‘Dark Side’ of Financial Innovation: The Cases
  of High Frequency Trading and Exchange Traded Funds}.
\newblock {\em International Journal of Entrepreneurship and Innovation
  Management}, {16}({1/2}):{51--72}, {2012}.

\bibitem{Bhattacharya2016}
Ayan Bhattacharya and Maureen O'Hara.
\newblock Can {ETF}s increase market fragility? {E}ffect of information
  linkages in {ETF} markets.
\newblock {\em Available at SSRN: http://ssrn.com/abstract=2740699}, 2016.

\bibitem{Malamud2015}
Semyon Malamud.
\newblock A dynamic equilibrium model of {ETF}s.
\newblock {\em Swiss Finance Institute Research Paper No. 15-37. Available at
  SSRN: http://ssrn.com/abstract=2662433 or
  http://dx.doi.org/10.2139/ssrn.2662433}, 2015.

\bibitem{da2013bellwether}
Zhi Da and Sophie Shive.
\newblock When the bellwether dances to noise: Evidence from exchange-traded
  funds.
\newblock {\em Available at SSRN 2158361}, 2013.

\bibitem{etfflavors}
{\em https://en.wikipedia.org/wiki/Exchange-traded\_fund\#Types AND
  http://etfdb.com/screener/}.

\bibitem{Axtell2001}
Robert~L. Axtell.
\newblock Zipf distribution of {U.S.} firm sizes.
\newblock {\em Science}, 293:1818--1820, 2001.

\bibitem{FamaandFrench1993}
Eugene.~F. Fama and R.~French Kenneth.
\newblock Common risk factors in the returns on stocks and bonds.
\newblock {\em Journal of Financial Economics}, 33:3--56, 1993.

\bibitem{MalPisSorUMPU11}
Y.~Malevergne, V.~Pisarenko, and D.~Sornette.
\newblock Testing the {P}areto against the lognormal distributions with the
  uniformly most powerful unbiased test applied to the distribution of cities.
\newblock {\em Physical Review E}, 83:036111, 2011.

\bibitem{Sornettebook04}
Didier Sornette.
\newblock Critical phenomena in natural sciences (chaos, fractals,
  self-organization and disorder: Concepts and tools).
\newblock {\em Springer Series in Synergetics, Heidelberg, 2nd ed.}, 2004.

\bibitem{Ramsden2000}
Jeremy Ramsden and Gy. Kiss-Haypál.
\newblock Company size distribution in different countries.
\newblock {\em Physica A: Statistical Mechanics and its Applications},
  277(1):220--227, 2000.

\bibitem{Axtell2006}
Robert Axtell.
\newblock Firm sizes: facts, formulae, fables and fantasies.
\newblock {\em \textit{in} Claudio Cioffi-Revilla, ed.: Power Laws in the
  Social Sciences (Cambridge University Press)}, 2006.

\bibitem{SimonandBonini1958}
Herbert~A. Simon and Charles~P. Bonini.
\newblock The size distribution of business firms.
\newblock {\em American Economic Review}, 46:607--617, 1958.

\bibitem{Marsili2005}
Orietta Marsili.
\newblock Technology and the size distribution of firms: Evidence from {D}utch
  manufacturing.
\newblock {\em Review of Industrial Organization}, 27:303--328, 2005.

\bibitem{Ijrl1977}
Yuji Ijri and Herbert~A. Simon.
\newblock Skew distributions and sizes of business firms.
\newblock {\em Studies in Mathematical and Managerial Economics (Book 24),
  (North- Holland, Amsterdam)}, 1977.

\bibitem{SaiMalSor09}
A.~Saichev, Y.~Malevergne, and D.~Sornette.
\newblock Theory of {Z}ipf's law and beyond.
\newblock {\em Lecture Notes in Economics and Mathematical Systems (Springer)},
  632, 2009.

\bibitem{LeraSor16}
Sandro~Claudio Lera and Didier Sornette.
\newblock Effects of mergers and acquisitions on firm size distributions.
\newblock {\em Swiss Finance Institute Research Paper}, (16-41), 2016.

\bibitem{MalSaiSor13}
Y.~Malevergne, A.~Saichev, and D.~Sornette.
\newblock {Z}ipf's law and maximum sustainable growth.
\newblock {\em Journal of Economic Dynamics and Control}, 37(6):1195--1212,
  2013.

\bibitem{scargle1982studies}
Jeffrey~D Scargle.
\newblock Studies in astronomical time series analysis. ii-statistical aspects
  of spectral analysis of unevenly spaced data.
\newblock {\em The Astrophysical Journal}, 263:835--853, 1982.

\bibitem{SornetteDSI98}
D.~Sornette.
\newblock Discrete scale invariance and complex dimensions.
\newblock {\em Physics Reports}, 297(5):239--270 (extended version at
  http://xxx.lanl.gov/abs/cond--mat/9707012), 1998.

\bibitem{ISI:000348403600019}
Grzegorz Wilk and Zbigniew Wlodarczyk.
\newblock {Tsallis Distribution Decorated with Log-Periodic Oscillation}.
\newblock {\em Entropy}, {17}({1}):{384--400}, {2015}.

\bibitem{ZhouSornetteLomb02}
W.-X. Zhou and D.~Sornette.
\newblock Statistical significance of periodicity and log-periodicity with
  heavy-tailed correlated noise.
\newblock {\em Int. J. Mod. Phys. C}, 13(2):137--170, 2002.

\bibitem{ZhouSornetteTurb02}
W.-X. Zhou and D.~Sornette.
\newblock Evidence of intermittent cascades from discrete hierarchical
  dissipation in turbulence.
\newblock {\em Physica D}, 165:94--125, 2002.

\bibitem{ZhouSornettePisTur03}
W.-X. Zhou, D.~Sornette, and V.~Pisarenko.
\newblock New evidence of discrete scale invariance in the energy dissipation
  of three-dimensional turbulence: Correlation approach and direct spectral
  detection.
\newblock {\em Int. J. Mod. Phys. C}, 14(4):459--470, 2003.

\bibitem{Huangetalarti00}
Y.~Huang, A.~Johansen, M.~W. Lee, H.~Saleur, and D.~Sornette.
\newblock Artifactual log-periodicity in finite-size data: Relevance for
  earthquake aftershocks,.
\newblock {\em J. Geophys. Res. (Solid Earth)}, 105:25451--25471, 2000.

\bibitem{Zhouetalfac3Dun05}
W.-X. Zhou, D.~Sornette, R.A. Hill, and R.I.M. Dunbar.
\newblock Discrete hierarchical organization of social group sizes.
\newblock {\em Proc. Royal Soc. London}, 272:439--444, 2005.

\bibitem{fuchs2014fractal}
Benedikt Fuchs, Didier Sornette, and Stefan Thurner.
\newblock Fractal multi-level organisation of human groups in a virtual world.
\newblock {\em Scientific reports}, 4, 2014.

\bibitem{zhou2002generalized}
Wei-Xing Zhou and Didier Sornette.
\newblock Generalized q analysis of log-periodicity: Applications to critical
  ruptures.
\newblock {\em Physical Review E}, 66(4):046111, 2002.

\bibitem{zhou2003nonparametric}
Wei-Xing Zhou and Didier Sornette.
\newblock Nonparametric analyses of log-periodic precursors to financial
  crashes.
\newblock {\em International Journal of Modern Physics C}, 14(08):1107--1125,
  2003.

\bibitem{GluzmanSornette02}
S.~Gluzman and D.~Sornette.
\newblock Log-periodic route to fractal functions.
\newblock {\em Phys. Rev. E}, 65:036142, 2002.

\bibitem{sornette1998linear}
Didier Sornette.
\newblock Linear stochastic dynamics with nonlinear fractal properties.
\newblock {\em Physica A: Statistical Mechanics and its Applications},
  250(1):295--314, 1998.

\bibitem{jogi1998fine}
Per J{\"o}gi, Didier Sornette, and Michael Blank.
\newblock Fine structure and complex exponents in power-law distributions from
  random maps.
\newblock {\em Physical Review E}, 57(1):120, 1998.

\bibitem{ouillon1996imaginary}
G~Ouillon, D~Sornette, A~Genter, and C~Castaing.
\newblock The imaginary part of rock jointing.
\newblock {\em Journal de Physique I}, 6(8):1127--1139, 1996.

\bibitem{ISI:A1988Q240100029}
C~Tsallis.
\newblock {Possible generalization of Boltzmann-Gibbs statistics}.
\newblock {\em {Journal of Statistical Physics}}, {52}({1-2}):{479--487}, {jul}
  {1988}.

\bibitem{Dragulescu2000}
Adrian Dragulescu and Victor~M. Yakovenko.
\newblock Statistical mechanics of money.
\newblock {\em Eur. Phys. J. B}, 17:723, (2000).

\bibitem{2005EL7070S}
D.~J.~B. {Soares}, C.~{Tsallis}, A.~M. {Mariz}, and L.~R. {da Silva}.
\newblock {Preferential attachment growth model and nonextensive statistical
  mechanics}.
\newblock {\em EPL (Europhysics Letters)}, 70:70--76, April 2005.

\bibitem{Cleymans:2015lxa}
J.~Cleymans and M.~D. Azmi.
\newblock {Large Transverse Momenta and Tsallis Thermodynamics}.
\newblock {\em J. Phys. Conf. Ser.}, 668(1):012050, 2016.

\bibitem{FamaandFrench2015}
Eugene.~F. Fama and R.~French Kenneth.
\newblock A five-factor asset pricing model.
\newblock {\em Journal of Financial Economics}, 116:1--22, 2015.

\bibitem{Piketty2014}
Thomas Piketty.
\newblock Capital in the twenty-first century.
\newblock {\em Harvard University Press}, 2014.

\bibitem{Campanale07}
Claudio Campanale.
\newblock Increasing returns to savings and wealth inequality.
\newblock {\em Review of Economic Dynamics}, 10:646--675, 2007.

\end{thebibliography}
\clearpage
\end{document}